\documentstyle[12pt]{article}
\newcommand{\be}{\begin{equation}}
\newcommand{\ee}{\end{equation}}
\newcommand{\ben}{\begin{enumerate}}
\newcommand{\een}{\end{enumerate}}

\newtheorem{Thm}{Theorem}[section]
\newtheorem{Prop}[Thm]{Proposition}
\newtheorem{Prob}[Thm]{Problem}
\newtheorem{Lem}[Thm]{Lemma}
\newtheorem{Rem}[Thm]{Remark}
\newcommand{\dowod}{\noindent{\bf Proof:} }
\newcommand{\qed}{\hspace{5.5in} Q.E.D. }

\newcommand{\ad}{\mbox{\rm ad}\, }
\newcommand{\Ldwa}{\,\stackrel{2}{\bigwedge}}
\newcommand{\w}{{\!}\wedge{\!}}

\font \msb=msbm10 scaled \magstep1

\newcommand{\rtimes}{\mbox{\msb o}\,}
\newcommand{\bR}{\mbox{\msb R} }
\newcommand{\bC}{\mbox{\msb C} }

\font \eul=eufm10 scaled \magstep2

\newcommand{\gotG}{\mbox{\eul g}}
\newcommand{\gotH}{\mbox{\eul h}}

\newcommand{\ar}{\alpha }
\newcommand{\tar}{\widetilde{\alpha} }
\newcommand{\br}{\beta }
\newcommand{\gr}{\gamma }
\newcommand{\dr}{\delta }
\newcommand{\Lr}{\Lambda }
\newcommand{\Om}{\Omega }
\newcommand{\trho}{\widetilde{\rho}}

\begin{document}

\title{\bf Poisson homogeneous spaces}
\author{{\bf S. Zakrzewski}  \\
\small{Department of Mathematical Methods in Physics,
University of Warsaw} \\ \small{Ho\.{z}a 74, 00-682 Warsaw,
Poland} }


\date{}
\maketitle

\begin{abstract}

General framework for Poisson homogeneous spaces of Poisson
groups is introduced. Poisson Minkowski spaces are discussed as
a particular example.

\end{abstract}

\section*{Introduction}
 Poisson Lie groups, introduced in \cite{D:ham} (see also
\cite{D,S-T-S,Lu-We,Lu}), allow one to describe  generalized
symmetries of classical mechanical systems and provide an
important tool for studying quantum deformations of Lie groups.
The classification of Poisson structures on a given group gives
usually a correct idea about the classification of the quantum
deformations of this group.

Together with Poisson (or quantum) groups, one has to consider
Poisson (or quantum) spaces on which these groups act. This, in
particular, rises the question of the classification of
homogeneous Poisson (or quantum) spaces for a given Poisson
(quantum) group.

In this paper we collect some fundamental facts concerning
Poisson homogeneous spaces and apply this to the classification
of Poisson Minkowski spaces. For each Poisson structure on the
Poincar\'{e} group (as studied recently in \cite{PPgr}) there is
exactly one Poisson structure on the Minkowski space, making it
a Poisson homogeneous space.

\section{Preliminaries}

We collect here basic definitions related to Poisson manifolds
\cite{We1} and Poisson groups
(main references: \cite{D:ham,D,S-T-S,Lu-We,Lu}).
We start with introducing a simple notation.

\vspace{1mm}

{\bf Notation}.

1. Let $M,N$ be manifolds and $f\colon M\to N$ a smooth map. For
any contravariant tensor $\xi$ at some point of $M$ we denote by
$f(\xi )$ the image of $\xi $ using $f$.  More precisely, if
$\xi\in \otimes^k T_xM$, then $f(\xi )$ will denote $(\otimes^k
T_xf)(\xi )$.

2. Let $M,N,P$ be manifolds and $f\colon M\times N\to P$ a
smooth map. Let $_xf\colon N\to P$, $f_y\colon M\to P$ be
defined by
\[
_xf (x) = f(x,y) = f_y (x) \qquad\qquad\mbox{{\rm
for}}\;\;x\in M, y\in N
\]
and suppose we denote $f$ as a product, $f(x,y) =xy$.
In this case we shall use the following notation:
\[
 \xi y := f_y (\xi ),\qquad\qquad x\eta := {_xf}(\eta ) ,
 \]
if $\xi$ is a contravariant tensor at some point of $M$, $\eta $
-- a contravariant tensor at some point of $N$.

A bivector field $\pi$ on a (smooth, real) manifold $M$ is said
to be  {\em Poisson}, if its associated bracket of functions,
\be \label{Pb}
\{f , g\}_{\pi } = <df\wedge dg ,\pi >
\ee
satisfies the Jacobi identity (in this case the bracket is said
to be {\em Poisson}). A bivector field $\pi $ is Poisson if and
only if the Schouten bracket $[\pi ,\pi ]$ is zero.  Poisson
bivectors are also called {\em Poisson structures}.  A {\em
Poisson manifold} is a pair $(M,\pi )$, where $M$ is a manifold
and $\pi $ is a Poisson structure on $M$.

Let $(M,\pi )$, $(N,\rho)$ be two Poisson manifolds.
A smooth map $f\colon M\to N$ is said to be a {\em Poisson
map} if $f(\pi (x)) = \rho (f(x))$  for each $x\in M$.

If $(M,\pi )$ is a Poisson manifold and $f\colon M\to N$ a
surjective submersion such that $f(\pi (x))=f(\pi (x'))$
whenever $f(x)=f(x')$, then there is exactly one Poisson
structure $\rho $ on $N$ such that $f$ is a Poisson map.
It is given by $\rho (f(x)):= f(\pi (x))$.

{\em Poisson product} $(M,\pi )\times (N,\rho )$ of two Poisson
manifolds $(M,\pi )$ and $(N,\rho )$ is the Poisson manifold
$(M\times N,\sigma )$ with the Poisson structure $\sigma $ given
by
\[
\sigma (x,y) :=
 \pi  (x)\oplus \rho (y)\in
 \Ldwa T_x M\oplus \Ldwa T_y N \subset \Ldwa
T_{(x,y)}(M\times N) \qquad \mbox{for}\;\; (x,y)\in
M\times N .
\]

A {\em Poisson group} is a Poisson manifold $(G,\pi )$ together
with a Poisson map $m\colon G\times G\to G$  (the product
Poisson structure on $G\times G$), such that $(G,m)$ is a group
(consequently, this group is a Lie group).

Since $m(X\oplus Y)= Xh + gY$ for $X\in T_gG$,
$Y\in T_hG$, the map $m$ is Poisson if and only if it is {\em
multiplicative}:
\be\label{multi}
\pi (gh) = \pi (g) h + g\pi (h) \qquad \mbox{for} \;\; g,h\in G,
\ee
where $gh:= m(g,h)$.

Let $(G,\pi )$ be a Poisson group. We denote by $\gotG$ the Lie
algebra of $G$. Since $\pi (e)= 0$, where $e$ is the group unit,
the bivector field $\pi $ has a well defined linearization at
$e$. The linearization of $\pi $ at $e$ will be denoted by $\dr
$. It is a linear map from $\gotG$ to $\Ldwa \gotG$, called {\em
cocommutator} or {\em cobracket}. The linear map
 $\dr \colon \gotG\to \Ldwa \gotG$ satisfies two conditions:

1. it is a 1-cocycle on $\gotG$
with values in $\Ldwa \gotG$ (with respect to the adjoint action)

2. it is a (linear) Poisson bivector on $\gotG$

Any pair $(\gotG ,\dr )$, where $\dr\colon \gotG\to \Ldwa \gotG$
is a linear map satisfying two above conditions, is said to be a
{\em Lie bialgebra}. The above mentioned correspondence between
Poisson groups and Lie bialgebras is known to be one to one, if
we consider only Poisson groups $(G,\pi )$ for which $G$ is
connected and simply connected.

\section{Poisson homogeneous spaces}

If $(G,\pi )$ is a Poisson group and $(M,\pi _M)$ is a Poisson
manifold then an action $\phi \colon G\times M\to M$ is said to
be a {\em Poisson action} if $\phi $ is a Poisson map. It is true
if and only if $\pi _M$ is $(G,\pi )$-{\em multiplicative} in
the following sense:
\be\label{multipa}
\pi _M(gx) = \pi (g) x + g\pi _M(x) \qquad \mbox{for} \;\; g\in
G, x\in M,
\ee
where $gx:= \phi (g,x)$. If the action $\phi $ is transitive then
$(M,\pi )$ is said to be a
 {\em Poisson homogeneous space}
of $(G,\pi )$.

In this section we prepare some tools to construct Poisson
homogeneous spaces of a given Poisson group. In particular, we
shall be interested in the following problem.
\begin{Prob} \label{problem}
 Given a Poisson group $(G,\pi )$ and a transitive action of $G$
on $M$, find all Poisson structures on $M$ such that the action
is Poisson.
\end{Prob}

We start with two simple observations:
\ben
 \item The difference of two $(G,\pi )$-multiplicative bivector
fields on $M$ is $G$-invariant. The sum of a $(G,\pi
)$-multiplicative bivector field and of a $G$-invariant bivector
field is $(G,\pi )$-multiplicative. (The space of $(G,\pi
)$-multiplicative bivector fields is an affine space modelled on
the space of $G$-invariant bivector fields.)
 \item  A $(G, \pi )$-multiplicative bivector field $\pi _M$ on
$M$ which vanishes at some point is automatically Poisson.
\een
The second observation follows from the fact that
if a $(G,\pi )$-multiplicative $\pi _M$ vanishes at $x_0\in M$,
then
\[
\pi _M(gx_0) = \pi (g)x_0
\]
hence $\pi _M$ is the image of $\pi$ under the orbital map
$g\mapsto gx_0$ (a surjective submersion).

In order to illustrate these facts, we consider first a special
case.  The Poisson group $(G,\pi )$ is fixed throughout this
section.

\subsection{Affine homogeneous space}

We consider $G$ acting on itself by left translations.
We are interested in $(G,\pi )$-multiplicative
Poisson bivector fields $\ar $ on $G$. Since $\pi $ is a
solution of this problem,  an arbitrary solution $\ar $ is a
Poisson bivector field which differs from $\pi $ by a
left-invariant field $g\mapsto A^l (g):=gA$, $A\in \Ldwa \gotG$,
hence
\be\label{affA}
\ar  = \pi  +  A^l,
\ee
for some $A\in \Ldwa \gotG$ (in fact, $A=\ar (e)$).
Any $\ar $ of the form (\ref{affA}) is $(G,\pi
)$-multiplicative. It is  Poisson if and only if
\be\label{calcul}
0=[\ar ,\ar ] = [\pi + A^l,\pi + A^l] = 2[\pi ,A^l] +[A^l,A^l].
\ee
It is easy to see that the right-hand side is always
left-invariant ($g\pi =\pi +B^r$, where $B^r$ is some
right-invariant field), hence it is zero if and only if
\be\label{drA}
2\dr (A) + [A,A] =0,
\ee
where $\dr (A) := [\pi ,A^l] (e)$, $[A,A] := [A^l,A^l](e)$
(since $\pi (e)=0$, $[\pi ,A^l](e)$ depends only on the value of
$A^l$ at $e$ and on the linearization $\dr$ of $\pi $).

We conclude that
$(G,\pi )$-multiplicative
Poisson bivector fields $\ar $ on $G$ are in 1--1 correspondence
with solutions $A$ of (\ref{drA}).

\begin{Rem} {\rm Poisson structures $\ar$ on $G$ such that}
\[
 \ar  - (\ar (e))^l = \pi
 \]
{\rm are exactly}  affine Poisson structures
{\rm on $G$ in the sense of \cite{D-S,Lu}, whose} associated
left multiplicative Poisson structure {\rm is} $\pi $. {\rm
Condition (\ref{drA}) has been derived in \cite{D-S,Lu}. It is
also shown in \cite{Lu}, that the solutions of (\ref{drA}) are in
1--1 correspondence with isotropic Lie subalgebras which are
complementary to $\gotG$ in the Manin triple Lie algebra $\gotG
\Join \gotG ^*$.}
\end{Rem}

{\em Special case}. \ It is easy to describe all
$(G,\pi )$-multiplicative Poisson bivector fields $\ar $ on $G$
which vanish at some point. Since $\ar (g) = \pi (g) +g\ar (e)$,
assuming $\ar (g_0)=0$ we obtain $\ar (e) = -g_0^{-1}\pi (g_0)$
and
\[
\ar (g) = \pi (g) -gg_0^{-1}\pi (g_0)= \pi
(gg_0^{-1}g_0)-gg_0^{-1}\pi (g_0) = \pi (gg_0^{-1})g_0 = (\pi
g_0)(g),
\]
hence $\ar $ is just the right translation of $\pi $ by $g_0$.
(The corresponding isotropic complementary Lie subalgebra in
the Manin triple is just the image of $\gotG ^*$ by the adjoint
action of $g_0$ in $\gotG\Join \gotG ^*$.)

\subsection{General situation}

We return to the situation of a general transitive action
$(g,x)\mapsto gx$ of $G$ on a manifold $M$.

First we formulate a result which allows to localize and linearize
the problem. The subgroup in $G$ stabilizing an element $x_0\in
M$ is denoted by $G_{x_0}$. The Lie algebra of $G_{x_0}$ is
denoted by $\gotG _{x_0}$.

\begin{Lem} For any $x_0\in M$ there is a 1--1 correspondence
between
\ben
\item $G$-multiplicative bivector fields $\pi _M$ on $M$

and
\item  elements $\rho\in \Ldwa T_{x_0}M $ such that
\be\label{muloc}
\rho = \pi (h)x_{0} + h\rho \qquad\qquad\mbox{for}\;\;h\in
G_{x_0}.
\ee
\een
The correspondence is given by:
\be\label{defrho}
\rho := \pi _M (x_0)
\ee
\be\label{defpi}
\pi _M(x) := \pi (g)x_0 + g\rho \qquad\mbox{where}\;\; g
\;\;\mbox{is any element of}\;\;G\;\;\mbox{such that}\;\;x
=gx_0.
\ee
\end{Lem}
\dowod If $\pi _M$ is $(G,\pi )$-multiplicative then $\rho =\pi
_M(x_0)$ satisfies (\ref{muloc}). Conversely, if $\rho$
satisfies (\ref{muloc}) then
 $\pi _M$ is well defined
by (\ref{defpi}), since
\[
 \pi (gh) x_0 + gh\rho = \pi (g)hx_0 + g\pi (h)x_0 + g(\rho -
\pi (h)x_0 ) = \pi (g)x_0 + g\rho
\]
for $g\in G$, $h\in G_{x_0}$. The $(G,\pi )$-multiplicativity
follows from
\[
\pi _M(gx)=\pi _M(gg_xx_0 )=\pi (gg_x)x_0 + gg_x\rho = \pi
(g)g_x x_0 + g(\pi (g_x)x_0 + g_x \rho ) = \pi (g)x  + g\pi _M
(x),
\]
where $g_x\in G$ is such that $x = g_xx_0$.

\qed

{\bf Corollary.} \  For connected $G_{x_0}$ we have a 1--1
correspondence between

\vspace{1mm}

 \ 1. \ $(G,\pi )$-multiplicative bivector fields $\pi _M$ on
$M$ and

 \ 2. \ elements $\rho\in \Ldwa T_{x_0}M $ such that
\be\label{mualg}
( \dr (X))x_0 + X\rho  =0 \qquad\qquad\mbox{for}\;\;X\in
\gotG _{x_0}.
\ee
 Denoting by $\trho$ any lift of $\rho$ to
$\Ldwa \gotG$,
\[
  \trho x_0 = \rho ,
	\]
we can write condition (\ref{mualg}) in the following equivalent
form:
\be\label{mulift}
 \dr (X) + \ad _X\trho \;\in \, \gotG\wedge \gotG _{x_0}
\qquad\qquad\mbox{for}\;\;X\in \gotG _{x_0}.
\ee

Now suppose $\rho$ satisfies (\ref{muloc}) hence the
corresponding $\pi _M$ defined by (\ref{defpi}) is $(G,\pi
)$-multi\-pli\-ca\-tive. Let $\trho\in\Ldwa \gotG$ be such that
$  \trho x_0 = \rho $. Since
\[
\pi _M (gx_0)=(\pi (g) +g\trho )x_0,
\]
$\pi _M$ is the image of $\pi +\trho ^l$ by the canonical
projection (the orbital map) from $G$ to $G/G_{x_0}=M$.
It is clear that
\[
[\pi _M ,\pi _M]=0 \qquad \Longleftrightarrow \qquad
[\pi +\trho ^l ,\pi +\trho ^l] x_0 =0.
\]
Since $[\pi +\trho ^l ,\pi +\trho ^l]$ is left-invariant (as in
formula (\ref{calcul})), the last equality is equivalent to
\be \label{drt}
(2\dr (\trho ) + [\trho ,\trho ])x_0 =0,
\ee
or,
\be \label{drt1}
2\dr (\trho ) + [\trho ,\trho ] \;\in \, \gotG\wedge
\gotG\wedge\gotG _{x_0} .
\ee

{\bf Conclusion:} \ solving problem~\ref{problem} is equivalent
to find $\rho$ satisfying (\ref{muloc}) (or (\ref{mulift}) if
$G_{x_0}$ is connected) and (\ref{drt}). ($\trho$ is any lift of
$\rho$ to $\Ldwa \gotG$.)

\subsection{The coboundary case} \label{cob}

In this subsection we suppose that $\dr$ is a coboundary of some
(classical $r$-matrix) $r\in \Ldwa \gotG$:
\[
\dr (X) = -\ad _X r.
\]
In this case, for $A\in \Ldwa\gotG$ we have
\[
\dr (A ) = [\pi , A^l ](e) = [r^{\rm right} -r^l,A^l](e)=
-[r,A],
\]
where $r^{\rm right}$ is the right-invariant field on $G$
corresponding to $r$. Conditions (\ref{mulift}) and (\ref{drt1})
take now the form
\be \label{1}
\ad _X (\trho - r) \;\in \, \gotG\wedge \gotG _{x_0}
\qquad\qquad\mbox{for}\;\;X\in \gotG _{x_0}
\ee
and
\be \label{2}
 ([\trho ,\trho ] - 2[r,\trho ])\;\in \, \gotG\wedge
\gotG\wedge\gotG _{x_0},
\ee
respectively.

\subsubsection{The case of the Poisson quotient}

Condition (\ref{mulift}) has as solution $\trho =0$ if and only
if
\be\label{coiso}
\dr (\gotG _{x_0})\subset \, \gotG\wedge \gotG _{x_0}.
\ee
This is exactly the case when $\pi $ is projectable by the
canonical projection from $G$ to $G/G_{x_0}=M$. The image of
$\pi $ by this projection is then the $(G,\pi)$-multiplicative
$\pi _M$ defined by $\pi _M(x_0)=\rho =0$.

The subgroup $G_{x_0}$ such that (\ref{coiso}) is satisfied is
called {\em coisotropic} in \cite{Lu}. Another way of expressing
condition (\ref{coiso}) is to require the annihilator
$\gotG_{x_0}^{\perp}$ of $\gotG_{x_0}$ to be a Lie subalgebra in
$\gotG ^*$ (with respect to the bracket defined by the dual map
of $\dr $).

A special case of this situation arises when $G_{x_0}$ is a
{\em Poisson subgroup} of $G$, i.e.
\[
\dr (\gotG_{x_0})\subset \, \gotG _{x_0}\wedge \gotG _{x_0}
\]
(equivalently: $\gotG _{x_0}^{\perp}$ is an ideal in $\gotG
^*$), cf. \cite{S-T-S,Lu-We}.

\vspace{1mm}

{\bf Example.} \ Recall that the `standard' Poisson structure
$\pi$ on a simple Lie group $G$ is defined in such a way that
the cobracket $\dr $ vanishes on some Cartan subalgebra
$\gotH\in \gotG$ (the corresponding Cartan subgroup $H$ is a
`classical' subgroup of the Poisson group $G$). It follows that
the quotient $G/H$ carries a natural Poisson structure (the
reduction of $\pi $ by the canonical projection) endowing $G/H$
with the structure of a Poisson $(G,\pi )$-homogeneous space.
In particular, taking $G=SU(2)$, $H=S^1\subset SU(2)$ we obtain
the `standard' Poisson sphere. Any Poisson structure on the
sphere which makes it a Poisson homogeneous $SU(2)$-space is a
sum of the standard Poisson structure and any $SU(2)$ invariant
bivector field on the sphere (in two dimensions all bivector
fields are Poisson), cf. \cite{Sheu}.

More general cases were described in \cite{Lu-We,KRR}.

\section{Poisson Minkowski spaces}

Poisson Poincar\'{e} groups in dimension $D=4$ have been
discussed and almost completely classified recently in
\cite{PPgr}. They are all coboundary.
The Lie algebra $\gotG$ has the semidirect structure
\[
\gotG = V\rtimes \gotH ,
\]
where $\gotH\cong sl (2,{\bC})$ is the Lorentz algebra and
$V$ is the subgroup of translations (which may be identified
with the Minkowski space).
The $r$-matrix has a decomposition
\[
 r= a\oplus b\oplus c \in \Ldwa V \oplus (V\wedge \gotH ) \oplus
\Ldwa \gotH .
\]
We have a simple proposition.
\begin{Prop}
For any classical $r$-matrix $r$ on the Poincar\'{e} Lie algebra
there is exactly one Poisson structure on $M:= G/H$ such that
the action of $G$ on $M$ is Poisson. This structure is
determined by the condition $\rho \equiv \pi _M(x_0) = a$.
\end{Prop}
\dowod
The uniqueness follows from the fact that zero is the only
$G$-invariant bivector field on $M$ (there are no non-zero
$\gotH$-invariants in $\Ldwa V$). We can always take the lift
$\trho$ belonging to $\Ldwa V$. Condition (\ref{1}) reads
\[
\ad _X (\trho - a -b -c) \;\in \, \gotG\wedge \gotH
\qquad\qquad\mbox{for}\;\;X\in \gotH,
\]
hence
\[
\ad _X (\trho - a) \;\in \, \gotG\wedge \gotH
\qquad\qquad\mbox{for}\;\;X\in \gotH,
\]
and therefore $\trho = a$. Since $[c,a]\in (\Ldwa V)\wedge
\gotH$ and $[b,a]=0$ (cf. \cite{PPgr}), we have
\[
 [a,a] - 2[r,a] \in (\Ldwa V)\wedge \gotH
 \]
and condition (\ref{2}) is satisfied.

\hfill $\Box $

\begin{Rem}\sloppy
The proposition has a straightforward generalization for the
$\bR ^{p+q}\rtimes O(p,q)$ groups discussed in ~\cite{PPgr}.
Next few remarks are also valid in more general context.
\end{Rem}

Now we derive a practical formula to compute  the Poisson
structure on $M$ corresponding to a given classical $r$-matrix
on $\gotG$. Due to the group law,
\[
(v,\Lr )(v',\Lr ') = (v+\Lr v',\Lr \Lr '),\qquad\qquad v,v'\in
V,\, \Lr ,\Lr '\in H ,
\]
the right and left translations of a vector $(u,X)\in \gotG
=V\rtimes \gotH$ by $g = (v,I)\equiv v\in V$ are given by
\be\label{transl}
 (u,X)v = (u+Xv,X),\qquad\qquad v(u,X)=(u,X).
 \ee
Using formula (\ref{defpi}) with $x_0=0$, $x=v=g$ and $\pi (g) =
rg-gr$, we obtain
\[
\pi (g) x_0 = \pi (v) x_0 = ((a+b+c)v-v(a+b+c))x_0=
(bv-vb)x_0 + (cv-vc)x_0,
\]
hence
\[
\pi _M(v) = (bv-vb)x_0 + (cv-vc)x_0 + a.
\]
Writing $b$ and $c$ in the form $ b= w_j\wedge Z_j$, $c =
X_k\wedge Y_k$ (summation convention) with $w_j\in V$,
$X_k,Y_k,Z_j\in \gotH$, we obtain (using (\ref{transl}))
\[
\pi _M(v) = a+ (w_j\wedge (Z_j v+ Z_j) - w_j\wedge Z_j ))x_0 +
((X_kv +X_k)\wedge (Y_kv+Y_k))x_0
\]
\be\label{piM}
\pi _M(v) = a + w_j\wedge Z_j v + X_k
v\wedge Y_k v = a + b(v) + c(v\otimes v),
\ee
with obvious notation. Therefore, $a$, $b$ and $c$ determine the
constant, linear and quadratic part of $\pi _M$, respectively.

It is interesting to note, that $a$, $b$ and $c$ (hence $r$) can
be recovered from $\pi _M$. Obviously, $a=\pi _M (0)$. In order
to see that $b$ and $c$ are determined by $\pi _M$, let us write
$b(v)$ and $c(v\otimes v)$ in more detail. Using the isomorphism
\[
 \Ldwa V \cong \gotH,\qquad \qquad x\wedge y \mapsto \Om
_{x,y}:= x\otimes \eta (y) - y\otimes \eta (x),
\]
where $\eta\colon V\to V^*$ is the metric tensor, we can think of
$b$ as an element of $V\otimes \Ldwa V\subset V\otimes V\otimes V$,
\[
 b = b^{\ar \br \gr}e_{\ar}\otimes e_{\br }\otimes
e_{\gr},\qquad\qquad b^{\ar\br\gr}=-b^{\ar\gr\br}.
\]
Here $e_{\ar }$ is a basis of $V$. We have
\[
b(v) = b^{\ar\br\gr } e_{\ar }\wedge (e_{\br }\eta
_{\gr\mu}-e_{\gr}\eta _{\br\mu})v^{\mu } = f^{\ar\br\gr}e_{\ar
}\wedge e_{\br} \eta _{gr\mu} v^{\mu},
\]
where
\[
 f^{\ar\br\gr} := b^{\ar\br\gr} - b^{\br\ar\gr}.
 \]
One can easily see that $b$ is computed from $f$ (hence from
$\pi _m$) as follows
\[
 b^{\ar\br\gr} = \frac12 (f^{\ar\br\gr} + f^{\gr\ar\br} -
f^{\br\gr\ar}).
\]
Similarly, we can think of $c\in \gotH \wedge \gotH$ as an
element of $\Ldwa V \wedge \Ldwa V\subset V\otimes V\otimes
V\otimes V$,
\[
c = c^{\ar\br\gr\dr} e_{\ar}\otimes e_{\br }\otimes
e_{\gr}\otimes e_{\dr } , \qquad\qquad c^{\ar\br\gr\dr} =
-c^{\br\ar\gr\dr} = -c^{\ar\br\dr\gr}= -c^{\gr\dr\ar\br}.
\]
For the symmetric bilinear form corresponding to $c(v\otimes v)$
we get the following expression:
\[
 \frac12 (c(v\otimes u)+c(u\otimes v))= h^{\ar\br\gr\dr}
e_{\ar}\wedge e_{\gr} \eta _{\br\mu}\eta _{\dr \nu} v^{\mu}
u^{\nu },
\]
where
\[
 h^{\ar\br\gr\dr} = 2(c^{\ar\br\gr\dr} + c^{\ar\dr\gr\br}).
 \]
It turns out that $c$ can be computed from $h$ (hence from $\pi
_M$):
\[
 c^{\ar\br\gr\dr} = \frac14 (h^{\ar\br\gr\dr} -
h^{\dr\gr\br\ar}).
\]

Below we  list the Poisson structures on $M$
corresponding to  the first six cases in the table of
$r$-matrices for the Poincar\'{e} group given  in
\cite{PPgr} (the case of non-vanishing $c$). Here
$x^0,x^1,x^2, x^3$ are the Lorentzian coordinates and $x^{\pm
}:= x^0 \pm x^3$. Not displayed brackets are zero.

\ben
 \item \ $r=\gr JH\w H + \ar e_+\w e_- + \tar e_1\w e_2 $
  \begin{eqnarray*}
 \{ x^1 ,x^{\pm}\} & = & \gr x^2 x^{\pm} \\
  \{ x^2 ,x^{\pm}\} & = & - \gr x^1 x^{\pm} \\
  \{ x^+ ,x^-\} & = & \ar \\
  \{ x^1 ,x^2\} & = & \tar
  \end{eqnarray*}
  Note that the two first equalities can be written in the form
  \[
	  \{ z ,x^{\pm}\}  =  -i \gr z x^{\pm} ,
		\]
  where $z:= x^1 + i x^2$.
 \item \ $r = JX_+\w X_+ +\br_1 (e_1\w X_+ - e_2\w JX_+ +  e_+\w H)
 + \br _2 e_+\w  JH $
   \begin{eqnarray*}
 \{ x^1 ,x^+\} & = & 2x^2 x^- + \br _1x^1 -\br _2 x^2 \\
  \{ x^2 ,x^+\} & = & - 2x^1 x^- + \br _1x^2  +\br _2 x^1 \\
  \{ x^+ ,x^-\} & = & - \br _1x^- \\
  \{ x^1 ,x^2\} & = & 4(x^- )^2
  \end{eqnarray*}
 \item \  $r = JX_+\w X_+ +\br (e_1\w X_+ - e_2\w
 JX_+ + e_+\w H) +\ar e_+ \w e_1 $
     \begin{eqnarray*}
 \{ x^1 ,x^+\} & = & 2x^2 x^- + \br x^1 -\ar  \\
  \{ x^2 ,x^+\} & = & - 2x^1 x^- + \br x^2  \\
  \{ x^+ ,x^-\} & = & - \br x^- \\
  \{ x^1 ,x^2\} & = & 4(x^- )^2
  \end{eqnarray*}
 \item \ $ r = JX_+\w X_+ + \br (e_1\w X_+ + e_2\w
 JX_+ +e_+ \w (\ar _1 e_1 + \ar _2 e_2)- \br ^2
e_1\w e_2 $
      \begin{eqnarray*}
 \{ x^1 ,x^+\} & = & 2x^2 x^- + \br x^1 -\ar _1 \\
  \{ x^2 ,x^+\} & = & - 2x^1 x^- - \br x^2 -\ar _2  \\
  \{ x^1 ,x^2\} & = & 4(x^- )^2 - \br ^2
  \end{eqnarray*}
 \item \  $r = H\w X_+ - JH\w JX_+ +  \gr JX_+\w X_+ $
      \begin{eqnarray*}
 \{ x^1 ,x^+\} & = & -2x^+ x^- -(x^2)^2 + 2 \gr x^2 x^- \\
  \{ x^2 ,x^+\} & = & x^1x^2- 2\gr x^1 x^-   \\
  \{ x^1 ,x^- \} & = & 2 (x^- )^2   \\
  \{ x^1 ,x^2\} & = & -2x^2x^- +4\gr (x^- )^2 \\
  \{ x^+,x^- \} & = & x^- x^1
  \end{eqnarray*}
\item \ $r = H\w X_+ +\br e_2\w X_+$
      \begin{eqnarray*}
 \{ x^1 ,x^+\} & = & -2x^+ x^- \\
  \{ x^2 ,x^+\} & = & \br x^1    \\
  \{ x^1 ,x^- \} & = & 2 (x^- )^2   \\
  \{ x^1 ,x^2\} & = & -2\br x^- \\
  \{ x^+,x^- \} & = & x^- x^1
  \end{eqnarray*}
\een

\begin{Rem}
$D=2$ Poincar\'{e} groups and corresponding Poisson Minkowski
spaces have been classified in \cite{poi}.
\end{Rem}

\end{document}